\documentclass[%
 aip,
 amsmath,amssymb,
 reprint,%
]{revtex4-1}

\usepackage{graphicx}
\usepackage{dcolumn}
\usepackage{bm}

\usepackage[utf8]{inputenc}
\usepackage[T1]{fontenc}
\usepackage{mathptmx}
\usepackage{etoolbox}
\usepackage{soul}
\usepackage{hyperref}
\hypersetup{
    colorlinks=true,
    linkcolor=blue,
    filecolor=magenta,      
    urlcolor=cyan,
    pdftitle={Overleaf Example},
    pdfpagemode=FullScreen,
    }

\usepackage{xcolor}

\newcommand{\VST}{$V_\mathrm{ST}\;$}
\newcommand{\VSB}{$V_\mathrm{SB}\;$}
\newcommand{\Vh}{$V_\mathrm{h}\;$}
\newcommand{\Vb}{$V_\mathrm{b}\;$}
\newcommand{\VES}{$V_\mathrm{ES}\;$}
\newcommand{\VG}{$V_\mathrm{G}\;$}
\usepackage{changes}
\setaddedmarkup{\textcolor{blue}{#1}}
\setdeletedmarkup{\textcolor{red}{\sout{#1}}}

\makeatletter
\def\@email#1#2{%
 \endgroup
 \patchcmd{\titleblock@produce}
  {\frontmatter@RRAPformat}
  {\frontmatter@RRAPformat{\produce@RRAP{*#1\href{mailto:#2}{#2}}}\frontmatter@RRAPformat}
  {}{}
}%
\makeatother
\begin{document}

\preprint{AIP/123-QED}

\title[Piezoresistive snap-through detection for bifurcation-based MEMS sensors]{Piezoresistive snap-through detection for bifurcation-based MEMS sensors
}
\author{I. Litvinov}
\affiliation{School of Mechanical Engineering, Tel Aviv University, Tel Aviv Jaffo, 6997801, Israel}
\email{ivanlitvinov@mail.tau.ac.il}
 
\author{G. Spaer Milo}%
\affiliation{School of Mechanical Engineering, Tel Aviv University, Tel Aviv Jaffo, 6997801, Israel}
\author{A. Liberzon}
\affiliation{School of Mechanical Engineering, Tel Aviv University, Tel Aviv Jaffo, 6997801, Israel}
\author{S. Krylov}
\affiliation{School of Mechanical Engineering, Tel Aviv University, Tel Aviv Jaffo, 6997801, Israel}

\date{\today}

\begin{abstract}
We report on the piezoresistive method for detecting stability loss events in  microelectromechanical (MEMS) sensors based on bifurcation. The method involves measuring the resistivity changes of an entire beam to detect snap-through transitions in an electrostatically actuated, bistable double-clamped crystalline Silicon (Si) microbeam.
The applicability of the suggested approach in two types of sensors - an ambient air temperature sensor or a mean air velocity sensor, is demonstrated. In both cases the bistable beam, serving as the sensing element,  is affected by the electrothermal Joule's heating and air cooling. The measured signal is obtained by monitoring the critical voltages of the snap-through transitions. Piezoresistive sensing is especially suitable for the response monitoring of the exposed to the environment, free-standing heated microbeam sensors, where optical, piezoelectric, or electrostatic interrogation methods are not applicable. The approach can be implemented in various bifurcation microsensors and for response monitoring of bistable actuators.  
\end{abstract}

\maketitle
Bistability, which refers to a mechanical structure's ability to maintain two stable equilibria when subjected to the same load, is a key feature in numerous microelectromechanical systems (MEMS).\cite{qiu2004curved, das2009pull, harne2013review} The increasing fascination with bistable devices is driven by a number of their unique advantageous characteristics. The bistability feature of curved double-clamped microbeams and curved microplates gives these devices exceptional sensitivity to external stimuli near the stability limits, making them ideal candidates for implementation in different types of sensors. The bistability is used in a wide variety of devices, including inertial switches~\cite{cao2019review, cao2020research}, logic gates~\cite{meng2021bistability}, accelerometers~\cite{halevy2020feasibility}, shock sensors~\cite{frangi2015threshold}, vibration energy harvesters~\cite{pellegrini2013bistable}, gas sensors~\cite{bouchaala2016nonlinear} and flow sensors~\cite{kessler2020flow}, to name a few.

Monitoring the mechanical response of bifurcation sensors, which is required to detect the bifurcation event, remains challenging.  Laser Doppler Vibrometry (LDV), widely used in laboratory experiments, cannot be implemented in future integrated sensors.~\cite{moran2013review, torteman2019electro} 
Commonly used in microsystems, electrostatic\cite{alattar2023tracking} interrogation requires positioning of electrodes close to the moving sensing element, which may not be suitable for implementation in devices exposed to the environment.~\cite{li2020recent} Also the electrostatic sensing is affected by radiation and cannot be used in an open space or in nuclear facilities.~\cite{shea2011effects} Piezoelectric sensing involves the deposition of multiple additional layers, usually highly stressed, which complicates the fabrication process and degrades performance. 

The piezoresistive effect has been known for considerable time and is widely used in many commercial and non–commercial applications due to its relatively simple and reliable nature.~\cite{tufte1962,fiorillo2018theory,barlian2009semiconductor, he2006giant}  The piezoresistive sensing, using a nanowire-type gauge, was recently implemented in a nano-electro-mechanical-system (NEMS)-based gyroscope (see~~\cite{dellea2017mems}). The use of piezoresistivity for the detection of  a microbeam resonator behavior was also reported.
~\cite{xiao2024dual,liu2024amplitude} In mechanical flow sensors, the piezoresistive effect is often used to detect deformations of cantilever-type structures caused by drag or lift forces.~\cite{svedin2003new, mathew2018review, zhang2010self, abolpour2022biomimetic} Despite the accumulation of literature on the application of bifurcation sensors~~\cite{halevy2020feasibility,frangi2015threshold}, there is a lack of discussion of the simple and convenient sensing approaches suitable for snap-through event detection. 

\begin{figure*}[!ht]
\includegraphics[width=1\textwidth]{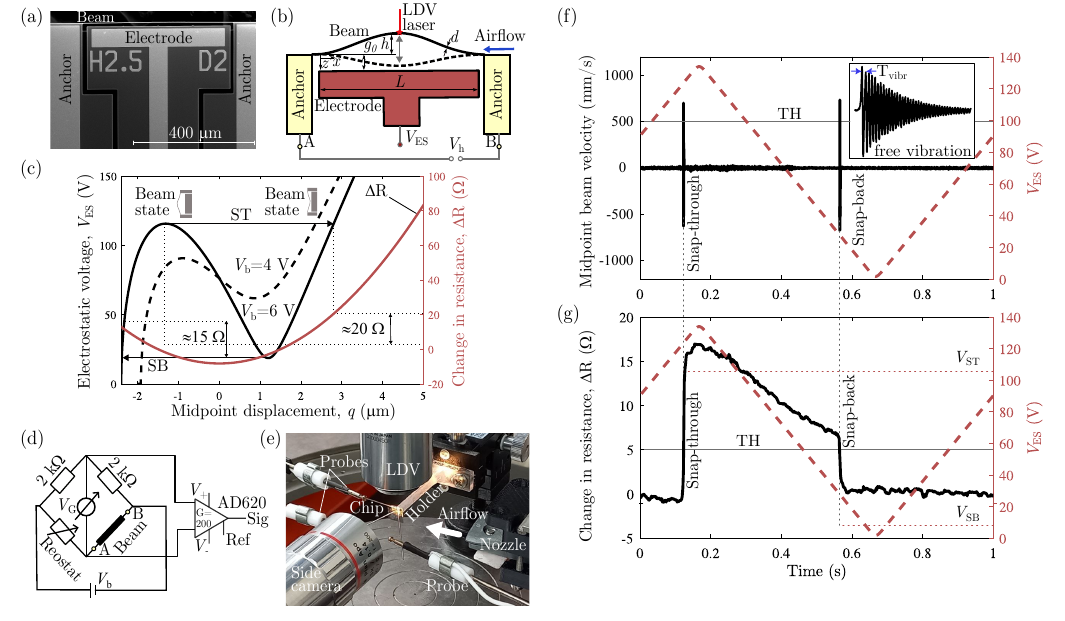}
\caption{ \label{fig:big} 
(a) Photo (Multimedia available online) and (b) schematic of bistable microbeam. (c) Model results: the limit point buckling curve - the actuating electrostatic voltage~\VES~as a function of the midpoint displacement of the beam $q$ and the piezoresistive change $\Delta \mathrm{R}$ of the beam resistance. The dashed and solid curves correspond to the Wheatstone bridge voltage $V_{\mathrm{b}}=$4~V and 6~V, respectively ($V_{\mathrm{h}}=V_{\mathrm{b}}/2$).  (d) The electrical circuit of the sensor with the Wheatstone bridge. (e) Photo of the experimental setup and sketch of the chip. The airflow direction is shown relative to the chip with the beam located at its edge. Experimental results: (f) The beam's midpoint velocity measured using LDV and the measured time history of the electrostatic voltage~\VES~cycle. (g)~The piezoresistive change $\Delta \mathrm{R}$ of the beam resistance detected by the Wheatstone bridge and the measured time history of the electrostatic voltage~\VES~cycle. The period of 1 s is shown. }
\end{figure*}

This paper presents a reliable method for detecting snap-through transitions in a double-clamped bistable microbeam, serving as a bifurcation-based sensing element. While the suggested approach can be implemented in a variety of bifurcation sensors, two examples of an air temperature sensor and an airflow velocity sensor are explored here. Robust detection of the snap-through transitions, using the piezoresistive effect of the entire microbeam, proved to be the most suitable and simple sensing approach. 
It is important to note that our unique architecture differs from conventional piezoresistively interrogated devices. In conventional devices, the size of the sensing guage is often defined by the selective implantation of dopants within a small area of the Si wafer, combined with a specially designed electrode configuration. However, in our architecture, the resistivity of the entire Si beam is measured, which significantly simplifies the design and improves the device's efficiency and scalability. 
We have carefully compared the microbeam snap-through buckling data obtained by the well-established LDV method and using the piezoresistivity. Moreover, the same heating voltage needed to assure bistability of the beam is also used for sensing purposes, thus demonstrating energy efficiency and feasibility of the suggested sensing scenario. 

An initially curved bell-shaped double-clamped microbeam of length $L$, width $b$, thickness $d$, and initial elevation of the midpoint $h$, is fabricated from a highly doped single crystal Si using the SOIMUMPs™ process [Fig.~\ref{fig:big}(a) (Multimedia available online) and Fig.~\ref{fig:big}(b)].~\cite{krylov2018micromechanical,kessler2020flow,kessler2021sampling,litvinov2024} The beams were fabricated from the device layer of the highly $(p++)$ Boron doped silicon on an insulator wafer, with resistivity $\rho_{e}=0.02\;\Omega \cdot \mathrm{cm}$ and (100) upper surface orientation. The beam axis is aligned with the $<110>$ crystallographic direction of Si.

The nominal (as designed) initial shape of the beam is defined by lithography and is described by the function $z_0(x) = -h \phi(x)$, where $\phi(x)=1/2(1-\cos{(2 \pi x /L)}) $ is the first buckling mode of a straight beam. The beam's dimensions are provided in Table~\ref{tab:dimensions}. The electrostatically actuated beam is designed to deflect in the in-plane (parallel to the substrate) $z$-direction. 
\begin{table}[!ht]
\small
\centering\caption{The nominal, as designed, parameters of the beam used in the experiments.}
\label{tab:dimensions}      
\begin{tabular}{lc}
\hline
Parameter & Size ($\mu$m)    \\ \hline
Length, $L$  & $500$  \\
Width, $b$ & $25$  \\ 
Thickness, $d$ & $2.0$  \\ 
Initial elevation, $h$  & $2.5$  \\
Beam-electrode distance, $g_0$  & $10$ \\ 
\hline
\end{tabular}
\end{table}

The operating principle of this sensor can be summarized as follows: the voltage~\VES~applied to the electrode [shown in Fig.~\ref{fig:big}(b)] influences the electrostatic force acting on the beam and causes the beam to be pulled toward the electrode. Exceeding a critical snap-through (ST) voltage~\VST~causes a dynamic transition from the first to the second stable equilibrium, as displayed in Fig.~\ref{fig:big}(c). The reduced-order model development is shown in the Supplementary Material. The beam undergoes snap-back (SB) when the voltage decreases below the lower threshold~\VSB~and the beam releases to its initial state. Applying the voltage $V_{\mathrm{h}}$ between the endpoints of the beam (points A and B) results in the Joule heating of the beam, an increase in the curvature of the beam, and a more significant split between the values of the critical voltages (\VST~and \VSB) values~\cite{litvinov2024}. 
The sensitivity of the bistable beam to changes in average beam temperature, which is in turn affected by the ambient air temperature and the air flow, increases, thereby making it more responsive to convective heat transfer to the surrounding air. Consequently, the bistable beam can be utilized to measure ambient air temperature through natural convection or air flow velocity through forced convection by detecting the values of \VST and \VSB.

The ability to reliably and accurately detect ST and SB events is critical for this type of a sensor's functionality and performance. The central objective of this research was to explore the feasibility and key features of the suggested robust snap-through and snap-back transition detection method.

The change in resistance of the entire microbeam due to transitions (ST and SB) was first estimated using the model. The transition between the equilibrium states caused a shift in the displacement of the beam and, as a consequence, a change in the electrical resistance of the beam, as shown in Fig.~\ref{fig:big}(c). In the framework of the single degree of freedom, reduced order (RO) model, the beam's mechanical response was parameterized by the midpoint displacement $q$. According to the model, the resistance change is $\approx 20\; \Omega$ and~$\approx 15\; \Omega$ for the ST and SB transitions, which could be easily detected in the experiment.

The experimental methodology implemented in the present work aligns with our previous study.~\cite{litvinov2024} The heating voltage~\Vh is applied between the A-B endpoints [see Fig.~\ref{fig:big} (b)] using the Wheatstone bridge to objectively discern the influence of the transition on the resistance of the beam. Under the condition that the bridge is balanced, a voltage of $V_{\mathrm{h}}=V_{\mathrm{b}}/2$ is applied to the beam [see the electrical circuit in~Fig.~\ref{fig:big}(d)]. To amplify the~\VG~indicator voltage from the Wheatstone bridge, the instrumental voltage amplifier (AD620) is used. To connect the device to the electrical circuit, needle probes are used and placed on the anchors of the beam and the electrode, with the aid of a video camera (IDS) and a horizontally mounted tube microscope (Navitar).

During the experiments, a ramp signal~\VES is applied using a Tektronix AFG3022C waveform generator and amplified by using a Trek PZD350A signal amplifier. The amplified voltage \VES is then used to apply an actuating voltage between 0 and 135~V to the electrode. Actuation loading and unloading cycles are performed to determine the values of~\VST and~\VSB using the piezoresistive effect. The National Instruments USB-6363 data acquisition system digitizes three signals: the voltage ramp signal (\VES), the signal obtained from the LDV system (Polytec), and the piezoresistive voltage signal (\VG). This enables the synchronization of the three signals. A 60-second signal is recorded at a sampling rate of 5 KHz. To collect statistics for 60 events of~\VST and~\VSB, a ramp frequency of 1 Hz is set, from which the mean and standard error are calculated.

To demonstrate the effectiveness of sensing two types of experiments we carried out. In the first ambient air temperature experiment, we used a climate control system in the room and measured the air temperature using the device synchronously with the CHY508BR thermometer.
In the second experiment, the chip with a horizontally oriented beam sensor was attached to a 50~mm long plastic 3D-printed chip extension holder in order to provide an undisturbed airflow along the beam, as shown in Fig.~\ref{fig:big}(e). The flow was in-plane, directed along the beam's axis and parallel to the chip. Airflow velocity is controlled by a pressure gauge (P-30PSIG-D / 5P, Alicat) that regulates the pressure of the air supply system. A rectangular flat nozzle (1" High Power, EXAIR) creates a wide, uniform rectangular jet flow directed at the microbeam from a distance of 40~mm. The air flow velocity is calibrated with a commercial anemometer (CTV 110, Kimo).

Figure~\ref{fig:big}(f) and Figure~\ref{fig:big}(g) illustrate the ramp-type electrostatic signal time history~\VES(t)~used to trigger the beam's bistable behavior. Figure~\ref{fig:big}(f) represents the midpoint velocity of the beam using LDV, showing two distinct peaks that correspond to ST and SB events. However, every peak is followed by an interval of free oscillations that decay in the beam with the vibration period $T_{\mathrm{vibr}}$, as shown in the zoom-in insert  (here with a sampling rate of 1 MHz). The free vibration frequency after ST/SB  was in good agreement with the fundamental frequency of the beam, approximately 110~kHz. The high speed at which the beam moves during ST causes the LDV system, operated in the velocity acquisition mode, to detect a spike in the time history in Fig.~\ref{fig:big}(f). We determine the critical events as the first and second peaks in the measured beam's velocity by the level-crossing method with the threshold of TH = 500 mm/s, marked as a horizontal solid gray line.

The piezoresistive signal of the microbeam stimulated by a ramp electrostatic voltage~\VES~ is shown in Fig.~\ref{fig:big}(g). The ST and SB jumps cause a rapid increase and decrease in the resistance of the entire beam. The resistance change at the SB transition is slightly lower than at the ST transition, which is close to the model prediction [Fig.~\ref{fig:big}(c)]. As shown in Fig.~\ref{fig:big}(g), ST and SB jumps can be easily detected using the change in resistance $\Delta$R based on~\VG signal monitoring. We determine critical events as the first and second jumps up and down in the~$\Delta$R~signal following crossing the threshold, marked as a solid horizontal gray line (TH = 5~$\Omega$ ). The corresponding \VST and \VSB values determined in two independent ways are shown as dotted lines in Fig.~\ref{fig:big}(g). Our results clearly show that the piezoresistivity signal is consistent with the LDV signal and can be used to detect ST and SB events.

\begin{figure}
\includegraphics[width=0.4\textwidth]{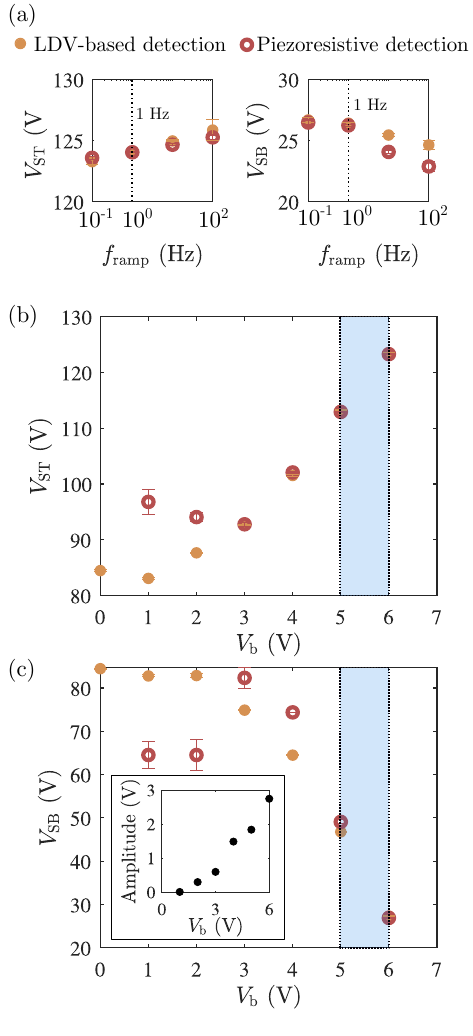}
\caption{\label{fig:heat} (a) The voltages of~\VST~and~\VSB~as a function of the ramp frequency $f_{\mathrm{ramp}}$. (b) Critical snap-through voltage~\VST~and (c) Critical snap-back voltage~\VSB~as a function of \Vb~ voltage. The shaded rectangle shows the limits of applicability of piezoresistivity-based sensing. The inset illustrates the dependence of the maximum amplitude of the piezoresistive signal on \Vb~levels.}
\end{figure}

To estimate the possible bandwidth of the sensor, the effect of the frequency of the electrostatic voltage ramp on the detection of ST\&SB is shown in Fig.~\ref{fig:heat}(a). The results based on LDV and piezoresistive data show that ST/SB voltages depend on the frequency of the electrostatic voltage ramp~\cite{kessler2021sampling}. Increasing the ramp frequency results in a mismatch between the data obtained from the piezoresistivity signal and the LDV-based data. This is likely due to the frequency limitations of the operational amplifier. Meanwhile, experimental findings indicate that raising the frequency beyond 600 Hz renders it unfeasible to reliably detect the quasistatic bistability behavior of a beam of this size.\cite{kessler2021sampling} The results in the following are shown at a ramp frequency of 1 Hz.

In our device, the same low current is used for the beam's heating (to tune the stability threshold) and for the piezoresistive sensing. This unique scenario simplifies fabrication, extends the design flexibility, and allows us to reduce the device's overall footprint. Therefore, the effect of beam overheating on detection capability based on monitoring \VST~and \VSB~voltages is to be considered. To achieve this, we varied the \Vb~voltage across the bridge from 0 to 6~V without flow and analyzed the functional relationship between~\VST~and~\VSB, as shown in Fig.~\ref{fig:heat}(b) and Fig.~\ref{fig:heat}(c). 
Our observations showed that as the heating voltage increased, there was a corresponding increase in~\VST~values and a decrease in~\VSB~ values, consistent with our previous research.~\cite{litvinov2024} Conversely, decreasing the bridge voltage $V_{\mathrm{b}}$~reduced the amplitude of the beam piezoresistivity signal in terms of the value \VG~[inset in Fig.~\ref{fig:heat}(c)].
However, a limitation of the piezoresistive method is reached for~\Vb<3~V, as the values of ST and SB voltage values obtained with this method began to diverge from those obtained with LDV. We expect that the values obtained with LDV remain valid. However, the limitation of the piezoresistive approach arises because the measurement is subject to significant uncertainty due to the higher noise level and the decreasing amplitude of the piezoresistive signal [inset in Fig.~\ref{fig:heat}(c)]. On the other hand, increasing the~\Vb~level above 7~V resulted in a latching effect, where the beam remains in the switched configuration at zero actuation voltage.
The SB is not achieved in this regime.~\cite{litvinov2024}

\begin{figure}[h]
\includegraphics[width=0.5\textwidth]{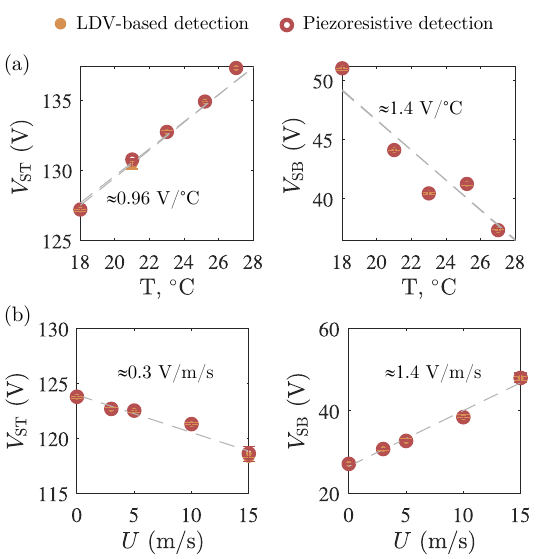}
\caption{ \label{fig:ST_SB} (a) Critical snap-through voltage \VST~and critical snap-back voltage \VSB~as a function of ambient air temperature T.
(b) Critical snap-through voltage \VST~and critical snap-back voltage \VSB~as a function of mean air velocity $U$. Dashed gray lines are linear trend lines. Error bars represent a standard error of the mean of each experimental point. }

\end{figure}

Figure~\ref{fig:ST_SB}(a) shows the results of the first experiment of~\VST~and~\VSB~as a function of ambient room temperature, without airflow from the nozzle.  Figure~\ref{fig:ST_SB}(b) shows the results of the air velocity experiment at constant air temperature (21~$^{\circ}$C) (note that we use two different sensors with two different microbeams).
In both experiments, we demonstrate a good agreement between the output signals of the ~\VST~and~\VSB, obtained with the piezoresistive sensing method of the entire beam and the LDV-based method. This indicates that the response detection using a piezoresistive signal is a feasible sensing method for free-standing bistable beam-based sensors. The bifurcation-based sensors are shown to be sensitive to changes in ambient air temperature and to airflow velocity. As previously shown~\cite{krylov2018micromechanical,kessler2020flow,litvinov2024}, the sensitivity to flow velocity at the voltage corresponding to the SB effect is much higher than the ST voltage values. This is a consequence of the nonlinear electrostatic force acting on the beam as it approaches the electrode. Indeed, the sensor's sensitivity is 0.96 V/$^{\circ}$C and 0.3 V/m/s for \VST~values and 1.4 V/$^{\circ}$C (V/m/s) for \VSB~values. Compared to the standard analog hot-wire micro anemometers~\cite{ejeian2019design}, our digital sensor offers significant advantages, including low power consumption and high sensitivity to the velocity (1.4~V/m/s). Theoretically, this device can extend the range of velocity measurements by higher \Vb levels. It is possible to design stiffer beams with higher initial elevation, use wider beams, or increase the overheating ratio by \Vb voltage and consequently the quality of the measurement signal.

We emphasize that detection is achieved by measuring the same electric current used for beam heating, controlled by the heating voltage \Vb between the beam’s ends. No additional layers are deposited on the sensing beam itself. This method enables direct detection of ST/SB using the bridge signal \VG, making the sensor suitable for field testing.

In summary, we present a simple, reliable and robust approach to accurately detect snap-through transitions, using a piezoresistive response of an entire bistable, double-clamped microbeam, applied as a bifurcation-based sensor. Our experiments demonstrate the sensor's sensitivity to surrounding air temperature through natural convection cooling of the microbeam or airflow velocity through forced convection. Overall, our findings suggest that this sensing method holds great promise for practical applications for bistable beams and, in general, bifurcation-based devices. In addition to its simplicity, one of the key advantages of the suggested approach is that it can be used for the interrogation of  downscaled nano-size beams, where common methods are not applicable.

See the supplementary material for details on the reduced-order model of the bistable beam.

The authors acknowledge the financial support of the Israel Innovation Authority, the NOFAR program, and Rafael Advanced Defense Systems Ltd. S. Krylov acknowledges the support of the Henry and Dinah Krongold Chair of Microelectronics.

\bibliography{library}
\end{document}